\input harvmac
\input amssym
\input epsf

\def\CE{{\cal E}}

\def\CN{{\cal N}}

\lref\AK{
  O.~Aharony and D.~Kutasov,
  ``Holographic duals of long open strings,''
Phys.\ Rev.\ D {\bf 78} (2008) 026005.
[arXiv:0803.3547 [hep-th]].
}
\lref\ASY{
  O.~Aharony, J.~Sonnenschein and S.~Yankielowicz,
  ``A holographic model of deconfinement and chiral symmetry restoration,''
Annals Phys.\  {\bf 322} (2007) 1420.
[hep-th/0604161].
}
\lref\AV{
  I.~Aref'eva and A.~Volovich,
  ``Composite $p$-branes in diverse dimensions,''
Class.\ Quant.\ Grav.\  {\bf 14} (1997) 2991.
[hep-th/9611026].
}
\lref\AVV{
  I.~Y.~Aref'eva, K.~S.~Viswanathan and I.~V.~Volovich,
  ``$p$-brane solutions in diverse dimensions,''
Phys.\ Rev.\ D {\bf 55} (1997) 4748.
[hep-th/9609225].
}
\lref\BLL{
  O.~Bergman, G.~Lifschytz and M.~Lippert,
  ``Holographic nuclear physics,''
JHEP {\bf 0711} (2007) 056.
[arXiv:0708.0326 [hep-th]].
}
\lref\BP{
  J.~L.~F.~Barbon and A.~Pasquinucci,
  ``Aspects of instanton dynamics in AdS/CFT duality,''
Phys.\ Lett.\ B {\bf 458} (1999) 288.
[hep-th/9904190].
}
\lref\BR{
  G.~E.~Brown and M.~Rho,
  ``Scaling effective Lagrangians in a dense medium,''
Phys.\ Rev.\ Lett.\  {\bf 66} (1991) 2720.
}
\lref\BSS{
  O.~Bergman, S.~Seki and J.~Sonnenschein,
  ``Quark mass and condensate in HQCD,''
JHEP {\bf 0712} (2007) 037.
[arXiv:0708.2839 [hep-th]].
}
\lref\Co{
  M.~S.~Costa,
  ``Composite M-branes,''
Nucl.\ Phys.\ B {\bf 490} (1997) 202.
[hep-th/9609181].
}
\lref\Coi{
  M.~S.~Costa,
  ``Black composite M-branes,''
Nucl.\ Phys.\ B {\bf 495} (1997) 195.
[hep-th/9610138].
}
\lref\CKP{
  R.~Casero, E.~Kiritsis and A.~Paredes,
  ``Chiral symmetry breaking as open string tachyon condensation,''
Nucl.\ Phys.\ B {\bf 787} (2007) 98.
[hep-th/0702155 [hep-th]].
}
\lref\CvTs{
  M.~Cvetic and A.~A.~Tseytlin,
  ``Nonextreme black holes from nonextreme intersecting M-branes,''
Nucl.\ Phys.\ B {\bf 478} (1996) 181.
[hep-th/9606033].
}
\lref\DN{
  A.~Dhar and P.~Nag,
  ``Tachyon condensation and quark mass in modified Sakai-Sugimoto model,''
Phys.\ Rev.\ D {\bf 78} (2008) 066021.
[arXiv:0804.4807 [hep-th]].
}
\lref\EV{
  M.~Edalati and J.~F.~Vazquez-Poritz,
  ``Chiral condensates in finite density holographic NJL model from string worldsheets,''
[arXiv:0906.5336 [hep-th]].
}
\lref\GKP{
  S.~S.~Gubser, I.~R.~Klebanov and A.~M.~Polyakov,
  ``Gauge theory correlators from noncritical string theory,''
Phys.\ Lett.\ B {\bf 428} (1998) 105.
[hep-th/9802109].
}
\lref\HHLY{
  K.~Hashimoto, T.~Hirayama, F.~-L.~Lin and H.~-U.~Yee,
  ``Quark mass deformation of holographic massless QCD,''
JHEP {\bf 0807} (2008) 089.
[arXiv:0803.4192 [hep-th]].
}
\lref\HHM{
  K.~Hashimoto, T.~Hirayama and A.~Miwa,
  ``Holographic QCD and pion mass,''
JHEP {\bf 0706} (2007) 020.
[hep-th/0703024 [hep-th]].
}
\lref\IMSY{
  N.~Itzhaki, J.~M.~Maldacena, J.~Sonnenschein and S.~Yankielowicz,
  ``Supergravity and the large N limit of theories with sixteen supercharges,''
Phys.\ Rev.\ D {\bf 58} (1998) 046004.
[hep-th/9802042].
}
\lref\KK{
  A.~Karch and E.~Katz,
  ``Adding flavor to AdS/CFT,''
JHEP {\bf 0206} (2002) 043.
[hep-th/0205236].
}
\lref\KMMW{
  M.~Kruczenski, D.~Mateos, R.~C.~Myers and D.~J.~Winters,
  ``Towards a holographic dual of large $N_c$ QCD,''
JHEP {\bf 0405} (2004) 041.
[hep-th/0311270].
}
\lref\KSZ{
  K.~-Y.~Kim, S.~-J.~Sin and I.~Zahed,
  ``Dense hadronic matter in holographic QCD,''
[hep-th/0608046].
}
\lref\KSZi{
  K.~-Y.~Kim, S.~-J.~Sin and I.~Zahed,
  ``The chiral model of Sakai-Sugimoto at finite baryon density,''
JHEP {\bf 0801} (2008) 002.
[arXiv:0708.1469 [hep-th]].
}
\lref\KSZii{
  K.~-Y.~Kim, S.~-J.~Sin and I.~Zahed,
  ``Dense holographic QCD in the Wigner-Seitz approximation,''
JHEP {\bf 0809} (2008) 001.
[arXiv:0712.1582 [hep-th]].
}
\lref\LT{
  H.~Liu and A.~A.~Tseytlin,
  ``D3-brane{--}D-instanton configuration and $\CN =4$ superYM theory in constant selfdual background,''
Nucl.\ Phys.\ B {\bf 553} (1999) 231.
[hep-th/9903091].
}
\lref\Ma{
  J.~M.~Maldacena,
  ``The large N limit of superconformal field theories and supergravity,''
Adv.\ Theor.\ Math.\ Phys.\  {\bf 2} (1998) 231.
[hep-th/9711200].
}
\lref\Mai{
  J.~M.~Maldacena,
  ``Wilson loops in large N field theories,''
Phys.\ Rev.\ Lett.\  {\bf 80} (1998) 4859.
[hep-th/9803002].
}
\lref\NSSY{
  S.~Nakamura, Y.~Seo, S.~-J.~Sin and K.~P.~Yogendran,
  ``A new phase at finite quark density from AdS/CFT,''
J.\ Korean Phys.\ Soc.\  {\bf 52} (2008) 1734.
[hep-th/0611021].
}
\lref\Se{
  S.~Seki,
  ``Intersecting D4-branes model of holographic QCD and tachyon condensation,''
JHEP {\bf 1007} (2010) 091.
[arXiv:1003.2971 [hep-th]].
}
\lref\SaSu{
  T.~Sakai and S.~Sugimoto,
  ``Low energy hadron physics in holographic QCD,''
Prog.\ Theor.\ Phys.\  {\bf 113} (2005) 843.
[hep-th/0412141].
}
\lref\SaSui{
  T.~Sakai and S.~Sugimoto,
  ``More on a holographic dual of QCD,''
Prog.\ Theor.\ Phys.\  {\bf 114} (2005) 1083.
[hep-th/0507073].
}
\lref\SekSi{
  S.~Seki and S.~-J.~Sin,
  ``Chiral condensate in holographic QCD with baryon density,''
JHEP {\bf 1208} (2012) 009.
[arXiv:1206.5897 [hep-th]].
}
\lref\SekSo{
  S.~Seki and J.~Sonnenschein,
  ``Comments on baryons in holographic QCD,''
JHEP {\bf 0901} (2009) 053.
[arXiv:0810.1633 [hep-th]].
}
\lref\SeoSi{
  Y.~Seo and S.~-J.~Sin,
  ``Baryon mass in medium with holographic QCD,''
JHEP {\bf 0804} (2008) 010.
[arXiv:0802.0568 [hep-th]].
}
\lref\Su{
  K.~Suzuki,
  ``D0-D4 system and QCD$_{3+1}$,''
Phys.\ Rev.\ D {\bf 63} (2001) 084011.
[hep-th/0001057].
}
\lref\Wi{
  E.~Witten,
  ``Anti-de Sitter space, thermal phase transition, and confinement in gauge theories,''
Adv.\ Theor.\ Math.\ Phys.\  {\bf 2} (1998) 505.
[hep-th/9803131].
}
\lref\Wii{
  E.~Witten,
  ``Baryons and branes in anti-de Sitter space,''
JHEP {\bf 9807} (1998) 006.
[hep-th/9805112].
}
\lref\Wiii{
  E.~Witten,
  ``Anti-de Sitter space and holography,''
Adv.\ Theor.\ Math.\ Phys.\  {\bf 2} (1998) 253.
[hep-th/9802150].
}
\lref\WXZ{
  C.~Wu, Z.~Xiao and D.~Zhou,
  ``Sakai-Sugimoto model in D0-D4 background,''
[arXiv:1304.2111 [hep-th]].
}

\Title{}
{\vbox{\centerline{A New Model of Holographic QCD and }
	\vskip12pt\centerline{ Chiral Condensate in Dense Matter}}}

\centerline{Shigenori Seki\footnote{$^\dagger$}{\tt sigenori@sogang.ac.kr}}
\medskip\centerline{\it Center for Quantum Spacetime (CQUeST)}
\centerline{\it Sogang University, Sinsu-dong, Mapo-gu, Seoul 121-741, Republic of Korea}
\bigskip
\centerline{Sang-Jin Sin\footnote{$^\ddagger$}
{\tt sjsin@hanyang.ac.kr}}
\medskip\centerline{\it Department of Physics, Hanyang University, Seoul 133-791, Republic of Korea}
 
\vskip .3in 
 
\centerline{\bf Abstract} 

We consider the model of holographic QCD with asymptotic freedom 
and gluon condensation in its vacuum. 
It consists of the color D4-branes and D0-branes as a background 
and the flavor D8-branes as a probe. 
By taking a specific field theory limit, the effective coupling decreases
to vanish in UV region. We then introduce the uniformly distributed baryons in terms of 
the baryon vertices  and  study the   density dependence of chiral condensate, 
which is evaluated using the worldsheet instanton method. 
In the confined phase, the chiral condensate 
as a function of density monotonically decreases in high baryon density. 
Such behavior is in agreement with the expectation, 
while in extremely low density it increases.  
We attribute this anomaly to the incorrect approximation of uniformity 
in very low density.  
In the deconfined phase the chiral condensate monotonically decreases 
in the whole region of density.

\Date{26 April 2013}

\newsec{Introduction}

Chiral symmetry  is one of the most important properties which control the behavior of 
the strongly interacting nucleon system  
and therefore studying  its order parameter, the chiral condensate, especially in the dense matter 
is the most  urgent and important to understand the properties of hadrons in nuclei, neutron stars 
and so on. So far many speculations have been made about the behavior 
of the chiral condensate in dense medium without definite proof. 
One of the intuitive and  acceptable ones 
is that the chiral condensate has a finite value 
at zero density and vanishes at a certain high density 
where the chiral symmetry is restored, and that it interpolates 
between the two by monotonically decreasing behavior \BR. 
However, no one succeeded in proving this  
from a first principle in a theory with strong coupling 
since there has not been any reliable calculational tools  
for the strongly interacting system especially in the presence of the dense fermions. 

Since the gauge/gravity correspondence \Ma\ is good 
for strongly interacting gauge theory,
it is tantalizing to see what happen 
if we apply this method to QCD and such application is called holographic QCD. 
The flavor physics of QCD has been studied in many models of 
holographic QCD, for example, 
the D3/D7  model \KK, the D4/D6  model \KMMW\ and 
the D4/D8  model \refs{\SaSu,\SaSui}. 
One can incorporate  baryons into those models 
by the use of the baryon vertices \Wii, the D5-branes wrapping $S^5$ in the D3-branes background 
or  the D4-branes wrapping $S^4$ in the D4-branes background. 
Such models with baryonic medium have been attracting many interests 
\refs{\BLL\KSZ\KSZi\KSZii\NSSY\SeoSi\EV{--}\SekSi}. 
However we have not found a model
which has the behavior of chiral condensate in complete agreement  
with the speculation mentioned above.

Although the D4/D8 model is one of the most successful models 
of holographic QCD, it has several shortcomings: 
current quark mass can not be incorporated, 
and as a consequence the chiral condensate can not be naturally introduced. 
Various prescriptions for this problem have been suggested 
by \refs{\CKP\HHM\BSS\AK\HHLY\DN{--}\Se}. 
The method by \AK\ allows us to calculate 
the quark mass and chiral condensate $\langle {\bar \psi}\psi \rangle$ from the open Wilson line, 
whose expectation value is given by the  minimal surface 
of the open string worldsheet surrounded by the D8-branes, 
the worldsheet instanton. 
Therefore the chiral condensate can be evaluated by the area 
of that minimal surface in the analogy of the computation of Wilson loop by \Mai. 
Following this method, we  have studied the chiral condensate 
in both confined and deconfined phases \SekSi. 
 There,   we found that  the chiral condensate 
decreases in low density but  increases in high density so that  
the low density behavior in this model 
agrees with the expectation, but the high density one does not, 
as in all other known models. 

The D4/D6 and D4/D8 models use the common background yielded by the D4-branes, 
in which the gluon sector of four-dimensional dual gauge theory 
becomes non-supersymmetric due to an anti-periodic $S^1$ compactification 
of fermions \Wi. 
This is an advantage of this background, 
because the ordinary QCD does not have supersymmetry. 
However the effective coupling $e^\phi$ diverges as one goes to UV, namely, 
the theory has opposite behavior to asymptotic freedom. 
In this paper we  improve this problem by introducing 
D0-branes smeared on the D4-branes.  We take a specific field theory limit to achieve the  asymptotic freedom.  
In general, the D$p$-branes smeared on the D$(p+4)$-branes play the role of 
instantons, so that the D$p$-branes provide a gluon condensate 
to the field theory on D$(p+4)$-branes \LT .

The rest of the  paper goes as follows. 
In Section 2 we suggest the new model whose background is given by 
taking the specific field theory limit of the supergravity solution 
of the D4-branes on which the D0-branes are smeared. 
In Section 3 we consider the D8-branes in the background of confined phase 
and then numerically calculate the chiral condensate 
depending on the density of baryons introduced 
as baryonic D4-branes (baryon vertices). 
In Section 4 we comment on the chiral condensate in deconfined phase which 
is analyzed in the similar way to the confined phase. 
Finally Section 5 is devoted to the conclusions. 

\newsec{A new model of holographic QCD by D4-branes and D0-branes}

In this section, we propose a new model of holographic QCD which consists of 
$N_4$ color D4-branes, $N_0$ D0-branes and $N_f$ flavor D8 and anti-D8-branes. 
The D0-branes are smeared on the color D4-branes and induce a gluon condensate. 
The theory has $U(N_4)$ color gauge symmetry  and $U(N_f) \times U(N_f)$ chiral symmetry. 
The flavor D-branes are treated as a probe embedded 
into the background produced by the color D-branes. 

\subsec{D4+D0 solution in IIA supergravity}

We consider $N_4$ D4-branes on which 
$N_0$ (Euclidean) D0-branes are homogeneously smeared. 
The configuration of these D-branes are shown in Table 1. 
\bigskip 
\centerline{\vbox{\offinterlineskip 
\halign{ 
\hskip2pt \hfil#\strut\hfil \hskip2pt 
& \hfil\vrule# \hskip1pc
& \hfil#\strut\hfil \hskip1pc
& \hfil#\strut\hfil \hskip1pc 
& \hfil#\strut\hfil \hskip1pc 
& \hfil#\strut\hfil \hskip1pc 
& \hfil#\strut\hfil \hskip1pc 
& \hfil#\strut\hfil \hskip1pc
& \hfil#\strut\hfil \hskip1pc 
& \hfil#\strut\hfil \hskip1pc 
& \hfil#\strut\hfil \hskip1pc 
& \hfil#\strut\hfil \hskip2pt \cr 
\noalign{\hrule} 
&& $X^0$ & $X^1$ & $X^2$ & $X^3$ & $X^4$ & $\rho$ & 6 & 7 & 8 & 9 \cr 
\noalign{\hrule \vskip1pt \hrule}
$N_4$ D4-branes && $\circ$ & $\circ$ & $\circ$ & $\circ$ & $\circ$ & & & & & \cr 
$N_0$ D0-branes && $\sim$ & $\sim$ & $\sim$ & $\sim$ & $\circ$ & & & & \cr 
\noalign{\hrule} 
} 
}}\nobreak 
\centerline{\vbox{\offinterlineskip 
\halign{ \strut# & #\hfil \cr
Table 1 & The configuration of the D4-branes and D0-branes. \cr
 & ``$\sim$'' stands for a smeared direction. \cr}}}
\medskip\noindent 
In IIA supergravity the non-extremal black brane solution describing such D-branes 
has already been known \refs{\CvTs\Co\AVV\Coi{--}\AV}. 
The solution has the following metric and dilaton:
\eqna\bbsolution
$$\eqalignno{
&ds^2 = \sqrt{H_0 \over H_4} dX_\mu dX^\mu 
	+{F \over \sqrt{H_0 H_4}} \bigl(dX^4\bigr)^2 
	+\sqrt{H_0 H_4}\biggl({d\rho^2 \over F} +\rho^2 d\Omega_4^2 \biggr) \,, &\bbsolution{a} \cr
&e^{\phi} = g_s \biggl({H_0^3 \over H_4}\biggr)^{1 \over 4} \,, \quad 
	g_s := e^{-\phi_0} \,, &\bbsolution{b}
}$$
where $\mu = 0,1,2,3$ 
which are contracted by $\eta_{\mu\nu} = {\rm diag}(-1,1,1,1)$.
$H_0$, $H_4$ and $F$ are the harmonic functions: 
\eqn\bbharmonic{
H_0(\rho) = 1 +\biggl({\rho_0 \over \rho}\biggr)^3 \,, \quad 
H_4(\rho) = 1 +\biggl({\rho_4 \over \rho}\biggr)^3 \,, \quad
F(\rho) = 1 -\biggl({\rho_h \over \rho}\biggr)^3 \,. 
}
$\rho_0$ and $\rho_4$ are related with the horizon, $\rho_h$, 
and the D-brane charges, 
$N_0$ and $N_4$, by
\eqn\bbhorcharge{
\rho_0^3 = -{1 \over 2}\rho_h^3 +\sqrt{{1 \over 4}\rho_h^6 +\pi^2\ell_s^6 g_s^2\biggl({(2\pi\ell_s)^4 N_0 \over V_4}\biggr)^2} \,, \quad 
\rho_4^3 = -{1 \over 2}\rho_h^3 +\sqrt{{1 \over 4}\rho_h^6 +\pi^2\ell_s^6 g_s^2 N_4^2} \,, 
}
where $V_4$ denotes the volume of the space on which the D0-branes are smeared, 
{\it i.e.}, $V_4 := \int dX^0 dX^1 dX^2 dX^3$. 

There are two Ramond-Ramond (RR) fluxes, $G_{(4)}$ and $G_{(8)}$, 
which are associated with the D4-branes and D0-branes respectively. 
Since the fluxes obey 
$(2\pi\ell_s)^{-3}\int G_{(4)} = N_4$ and 
$(2\pi\ell_s)^{-7}\int G_{(8)} = N_0$ by the definition of the D-brane charges, 
one can write down the fluxes as 
\eqn\bbRRflux{
G_{(4)} = {(2\pi\ell_s)^3 N_4 \over \Omega_4} d\Omega_4 \,, \quad
G_{(8)} = {(2\pi\ell_s)^7 N_0 \over V_4 \Omega_4} dX^0 \wedge dX^1 \wedge dX^2 \wedge dX^3 \wedge d\Omega_4 \,, 
}
where $\Omega_4$ is the volume of unit four-sphere, {\it i.e.}, 
$\Omega_4 = 8\pi^2/3$. 

\subsec{The field theory limit}

Let us consider the Maldacena limit \IMSY, 
\eqn\Maldlim{
{\rho \over \ell_s^2} =: U\ \hbox{is fixed}, \quad 
\ell_s \to 0 \,.
}
Since $U$ has the dimension of energy, 
the system under the limit \Maldlim\ lies in the finite energy configuration. 
The leading term of $H_4$ as $\ell_s \to 0$ is 
\eqn\redHfour{
H_4 = {g_{\rm YM5}^2 N_4 \over 2 \ell_s^4 U^3} + \CO(\ell_s^0) \,,
}
where $g_{\rm YM5}$ is the five-dimensional Yang-Mills coupling 
defined by $g_{\rm YM5}^2 = 2\pi g_s \ell_s$, 
while $H_0$ is 
\eqn\hzeroUco{
H_0 = 1 -{1 \over 2}\biggl({U_h \over U}\biggr)^3 
	+\sqrt{{1 \over 4}\biggl({U_h \over U}\biggr)^6 +{1 \over U^6}{(2\pi)^8 g_{\rm YM5}^4n_0^2 \over 4}} \,,
}
where $U_h := \rho_h \ell_s^{-2}$. 
We fix the density of D0-branes,
\eqn\dzerodens{
n_0 := {N_0 \over V_4} \,,
}
to be large but finite, and introduce the UV cutoff, $U_\infty$, by 
\eqn\Uuvcutoff{
(U_h^3 \leq)\  U^3 \leq U_\infty^3 \ll {(2\pi)^4 g_{\rm YM5}^2 n_0 \over 2} \,,
}
so that we concentrate on the IR region defined by \Uuvcutoff. 
Actually the analyses in the rest of this paper basically depend 
on the configuration of IR. 
Thus the harmonic function $H_0$ is reduced to\foot{Without focusing on the IR region, all the terms in \hzeroUco\ survive. 
Such a model was studied by \refs{\BP\Su{--}\WXZ}.  
}
\eqn\redHzero{
H_0 = {(2\pi)^4 g_{\rm YM5}^2 n_0 \over 2 U^3} \,. 
}
Substituting \redHfour\ and \redHzero\ into \bbsolution{}, we obtain
\eqn\preconfmet{\eqalign{
&\ell_s^{-2} ds^2 = (2\pi)^2 \sqrt{n_0 \over N_4}dX_\mu dX^\mu
	+{2 U^3 \over (2\pi)2 g_{\rm YM5}^2 N_4}\sqrt{N_4 \over n_0} F_U \bigl(dX^4\bigr)^2 \cr
&\qquad\qquad	+{(2\pi)^2 g_{\rm YM5}^2 N_4 \over 2 U^3}\sqrt{n_0 \over N_4}\biggl({dU^2 \over F_U} +U^2 d\Omega_4^2\biggr) \,, \cr
&e^{\phi} = (2\pi)^2 \biggl({n_0 \over N_4}\biggr)^{3 \over 4} \sqrt{g_{\rm YM5}^6 N_4 \over 2 U^3} \,, \quad
F_U = 1- \biggl({U_h \over U}\biggr)^3 \,. 
}}
The remarkable point is that, within this region, 
the coupling $e^\phi$ decreases as a function of $U$, 
which is interpreted as an energy scale. 
The RR fluxes \bbRRflux\ does not change by our field theory limit. 
Note that taking $n_0 \to 0$ limit is not consistent with \Uuvcutoff.

For later convenience, 
we introduce the coordinates rescaled by $R$ and $\ell_s$ as 
\eqn\resdimlesscor{
x^\mu := {X^\mu \over R} \,, \quad 
y := {X^4 \over R} \,, \quad 
r := {\ell_s^2 U \over R} \,, \quad 
R^3 := \pi g_s \ell_s^3 N_4 \,, 
}
so that $x^\mu$, $y$ and $r$ are dimensionless coordinates. 
We defined the radius $R$ by following the notation in \refs{\KMMW,\SaSu}. 
Then the metric and dilaton \preconfmet{} become 
\eqna\CONFmetricdl
$$\eqalignno{
&R^{-2} ds^2 = 
	\sqrt{\nu} dx_\mu dx^\mu
	+{r^3 f_M(r) \over \sqrt{\nu}} dy^2 
	+{\sqrt{\nu} \over r^3} \biggl({dr^2 \over f_M(r)} +r^2 d\Omega_4^2 \biggr) \,, &\CONFmetricdl{a} \cr
&e^{\phi} = g_s {\nu^{3/4} \over r^{3/2}} \,, \quad 
f_M(r) = 1 -\biggl({r_M \over r}\biggr)^3 \,, \quad 
\nu := {(2\pi\ell_s)^4 n_0 \over N_4} \,, \quad 
v_4 := \int d^4x = {V_4 \over R^4} \,, \cr
&&\CONFmetricdl{b} \cr
}$$
where $r_M := \ell_s^2 U_h /R$. 
The factor, $\ell_s^4$, in $\nu$ is an artifact 
caused by \resdimlesscor, 
and we recall that, in taking the field theory limit, 
$n_0$ is finite but $\nu$ is not. 
The RR fluxes \bbRRflux\ are rewritten as 
\eqn\CONFRRfluxdl{
G_{(4)} = {(2\pi\ell_s)^3 N_4 \over \Omega_4} d\Omega_4 \,, \quad
G_{(8)} = {(2\pi\ell_s)^3 R^4 N_4\nu \over \Omega_4} dx^0 \wedge dx^1 \wedge dx^2 \wedge dx^3 \wedge d\Omega_4 \,.
} 

We compactify the $y$ direction by a circle with a period $\beta_y$ 
in the same manner as \Wi, 
so that we obtain the four-dimensional dual gauge theory. 
In order for the smoothness at $r = r_M$, 
the period $\beta_y$ has to be 
$$
\beta_y = {4\pi \sqrt{\nu} \over 3r_M^2} \,.
$$
Then the Kaluza-Klein mass 
and the Yang-Mills coupling in the four-dimensional dual gauge theory 
are described as 
\eqn\MkkgYM{\eqalign{
M_{\rm KK} &:= {2\pi \over R \beta_y} = {3r_M^2 \over 2\sqrt{\nu}R} 
={3U_h^2 \over 4\pi^2 g_{\rm YM5}^2 N_4} \sqrt{N_4 \over n_0} \,, \cr 
g_{\rm YM}^2 &:= 2\pi g_s \ell_s M_{\rm KK} = g_{\rm YM5}^2 M_{\rm KK} \,.
}}

The background given by \CONFmetricdl{} and \CONFRRfluxdl\ corresponds to the confined phase of the dual gauge theory. 
If we take the T-dual on the $y$ direction, 
this D4+D0 becomes the D3+D($-1$) system which was studied by \LT. 
The D$(-1)$-brane plays a role of axion/gluon condensation in the dual gauge theory, 
because a D$p$-brane can be realized as an instanton on D$(p+4)$-branes.
Therefore the D0-branes smeared on the D4-branes behave as the instantons 
and provide the ``axion condensation''.

We need to clarify the parameter region in which 
the supergravity approximation is valid. 
First, in order to suppress the loop correction, 
the effective string coupling must be small, namely, $e^\phi \ll 1$. 
Since $e^\phi$ decreases as $r$, 
this constraint is realized as $g_s \nu^{3/4}r_M^{-3/2} \ll 1$, that is, 
\eqn\sgrCSTone{
{\lambda \over N_4^{4/3}} \sqrt{n_0 \over N_4} \ll  M_{\rm KK}^2 \,, 
}
where $\lambda$ is the 't Hooft coupling, $\lambda := g_{\rm YM}^2 N_4$, 
in four dimensions. 
Second, the curvature must be smaller than $\ell_s^{-2}$ for neglecting 
a higher curvature correction. 
We calculate the scalar curvature 
from the metric \CONFmetricdl{a},
\eqn\scurvature{
{\cal R}(r) = {9 \over \sqrt{\nu}R^2 r^2}(r_M^3 + r^3) \,. 
}
Since this is the increasing function for large $r$, 
the UV cutoff $r_\infty (= \ell_s^2 U_\infty/R)$ is manifestly necessary\foot{In our 
field theory limit, the harmonic function $H_0$ in \bbharmonic\ 
is reduced to $H_0 = \nu/r^3$. 
Let us naively imagine another limit where $H_0$ has the form, 
$H_0 = 1 +\nu/r^3$ ({\it cf}.~\refs{\BP,\Su}). 
The scalar curvature in this case is
$$
{\cal R}(r) = {9 \over \sqrt{\nu} R^2r^2(1+r^3/\nu)^{5/2}}
\left[ r_M^3 +r^3 -{r^3 \over \nu}(r^3 -2r_M^3) -{r^6 \over 4\nu^2}(r^3 -r_M^3)\right] \,.
$$
At small $r$, this curvature coincides with \scurvature. 
In other words, our theory focuses on the small $r$ region in 
the theory of $H_0 = 1 +\nu/r^3$. 
} 
so that ${\cal R}(r_\infty) \ll \ell_s^{-2}$, namely, 
$r_\infty \ll \sqrt{\nu}R^2 \ell_s^{-2}$.
This constraint with $r_M < r_\infty$ leads to  
\eqn\sgrCSTtwo{
M_{\rm KK}^2 \ll \lambda \sqrt{n_0 \over N_4} \,.
}

Here we recall \Uuvcutoff, which provides the condition, 
$r_M \ (\leq r_\infty) \ll \nu^{1/3}$. 
This is equivalent in terms of the field theory parameters to 
\eqn\uvccond{
M_{\rm KK}^2 \ll {1 \over \lambda}\sqrt{n_0 \over N_4} \,.  
}
The three conditions, \sgrCSTone, \sgrCSTtwo\ and \uvccond, are 
consistent with the 't Hooft limit, namely, 
the $N_4 \to \infty$ limit with fixing $\lambda$ to a large finite number 
as well as fixing the ratio, $n_0/N_4$, to a very large finite number.

We can also write down the background geometry 
corresponding to the deconfined phase. 
By exchanging the coordinates and double Wick-rotation, namely, 
$(x^0, y) \to (iy, ix^0)$, 
in \CONFmetricdl{} and \CONFRRfluxdl, 
we obtain the metric, 
\eqna\DECmetricdl
$$\eqalignno{
&R^{-2} ds^2 = -{r^3 f_T(r) \over \sqrt{\nu}} \bigl(dx^0\bigr)^2
	+\sqrt{\nu} \bigl(d{\vec x}^2 +dy^2 \bigr)  
	+{\sqrt{\nu} \over r^3} \biggl({dr^2 \over f_T(r)} +r^2 d\Omega_4^2 \biggr) \,, &\DECmetricdl{a} \cr
&e^{\phi} = g_s {\nu^{3/4} \over r^{3/2}} \,, \quad 
f_T(r) = 1 -\biggl({r_T \over r}\biggr)^3 \,, \quad 
\nu = {(2\pi\ell_s)^4 N_0 \over R^4 v_4}{1 \over N_4} \,, \quad 
v_4 := \int dx^1 dx^2 dx^3 dy \,, \cr 
&&\DECmetricdl{b} \cr
}$$
and the RR fluxes, 
\eqn\DECRRfluxdl{
G_{(4)} = {(2\pi\ell_s)^3 N_4 \over \Omega_4} d\Omega_4 \,, \quad
G_{(8)} = {(2\pi\ell_s)^3 R^4 N_4\nu \over \Omega_4} dx^1 \wedge dx^2 \wedge dx^3 \wedge dy \wedge d\Omega_4 \,.
} 
The transition between 
the background geometry for the confined phase \CONFmetricdl{} 
and the one for the deconfined phase \DECmetricdl{} 
is known as the Hawking-Page transition. 
Since we exchanged the time direction with the radial one, 
the configuration of D-branes is slightly modified as Table 2.  
\bigskip 
\centerline{\vbox{\offinterlineskip 
\halign{ 
\hskip2pt \hfil#\strut\hfil \hskip2pt 
& \hfil\vrule# \hskip1pc
& \hfil#\strut\hfil \hskip1pc
& \hfil#\strut\hfil \hskip1pc 
& \hfil#\strut\hfil \hskip1pc 
& \hfil#\strut\hfil \hskip1pc 
& \hfil#\strut\hfil \hskip1pc 
& \hfil#\strut\hfil \hskip1pc
& \hfil#\strut\hfil \hskip1pc 
& \hfil#\strut\hfil \hskip1pc 
& \hfil#\strut\hfil \hskip1pc 
& \hfil#\strut\hfil \hskip2pt \cr 
\noalign{\hrule} 
&& $x^0$ & $x^1$ & $x^2$ & $x^3$ & $y$ & $r$ & 6 & 7 & 8 & 9 \cr 
\noalign{\hrule \vskip1pt \hrule}
$N_4$ D4-branes && $\circ$ & $\circ$ & $\circ$ & $\circ$ & $\circ$ & & & & & \cr 
$N_0$ D0-branes && $\circ$ & $\sim$ & $\sim$ & $\sim$ & $\sim$ & & & & \cr 
\noalign{\hrule} 
} 
}}\nobreak 
\centerline{\vbox{\offinterlineskip 
\halign{ \strut# & #\hfil \cr
Table 2 & The configuration of the D4-branes and D0-branes for the deconfined phase. \cr}}}
\medskip\noindent 
Since the geometry \DECmetricdl{} is a black hole, 
we can introduce a temperature 
by compactifying the Euclidean time direction $\tau\, (:=ix^0)$ with a period $\beta_\tau$ 
which is determined to be $(4\pi/3)\sqrt{\nu} r_T^{-2}$. 
Then it leads to the (Hawking) temperature, 
\eqn\bbtemp{
T = {1 \over R \beta_\tau} = {3 r_T^2 \over 4\pi \sqrt{\nu} R} \,. 
}

It is remarkable that the power of effective couplings $e^\phi$, 
\CONFmetricdl{b} and \DECmetricdl{b}, 
has a negative value, $-3/2$. 
This implies the UV asymptotic freedom. 
On the other hand, the dilaton in the D4-branes background without D0-branes 
behaves like $e^\phi \sim r^{+3/4}$ \refs{\KMMW\SaSu{--}\SaSui}, 
that is, asymptotically non-free. 
Therefore we expect that the UV behavior in our holographic QCD model is 
improved by virtue of the D0-branes.

\newsec{Dense baryons in the confined phase} 

\subsec{The action of flavor D8-branes}

In order to consider the flavor dynamics in the confined phase, 
we embed $N_f$ D8-branes and $N_f$ anti-D8-branes 
into the background \CONFmetricdl{}, in which 
the D8-branes and anti-D8-branes are connected with each other, 
so that they become $N_f$ U-shape D8-branes (see fig.~1). 
This implies the chiral symmetry breaking from $U(N_f) \times U(N_f)$
to $U(N_f)$. 
\bigskip\vbox{
\centerline{\epsfbox{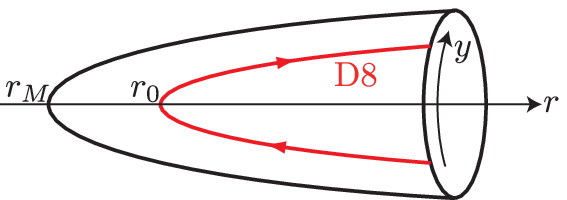}} \smallskip
\centerline{\vbox{\offinterlineskip 
\halign{ \strut# & #\hfil \cr
\fig\figphases{} & The embedding of D8-branes.  \cr
}}}
}\bigskip\nobreak
The D8-branes are stretched on $(x^0,x^1,x^2,x^3,r,S^4)$ 
and the collective coordinate $y$ depends only on $r$ (see Table 3). 
\bigskip 
\centerline{\vbox{\offinterlineskip 
\halign{ 
\hskip2pt \hfil#\strut\hfil \hskip2pt 
& \hfil\vrule# \hskip1pc
& \hfil#\strut\hfil \hskip1pc
& \hfil#\strut\hfil \hskip1pc 
& \hfil#\strut\hfil \hskip1pc 
& \hfil#\strut\hfil \hskip1pc 
& \hfil#\strut\hfil \hskip1pc 
& \hfil#\strut\hfil \hskip1pc
& \hfil#\strut\hfil \hskip1pc 
& \hfil#\strut\hfil \hskip1pc 
& \hfil#\strut\hfil \hskip1pc 
& \hfil#\strut\hfil \hskip2pt \cr 
\noalign{\hrule} 
&& $x^0$ & $x^1$ & $x^2$ & $x^3$ & $y$ & $r$ & 6 & 7 & 8 & 9 \cr 
\noalign{\hrule \vskip1pt \hrule}
$N_4$ D4-branes && $\circ$ & $\circ$ & $\circ$ & $\circ$ & $\circ$ & & & & & \cr 
$N_0$ D0-branes && $\sim$ & $\sim$ & $\sim$ & $\sim$ & $\circ$ & & & & \cr 
\noalign{\hrule} 
$N_f$ D8-branes && $\circ$ & $\circ$ & $\circ$ & $\circ$ & $y(r)$ & $\circ$ & $\circ$ & $\circ$ & $\circ$ & $\circ$ \cr 
\noalign{\hrule}
} 
}}\nobreak 
\centerline{\vbox{\offinterlineskip 
\halign{ \strut# & #\hfil \cr
Table 3 & The embedding of the flavor D8-branes. \cr}}}
\medskip\noindent 
Then the action of D8-branes without a worldvolume gauge field is 
\eqn\conActzero{
S_8 = -\CT\int d^4xdr\, \nu\sqrt{r^2f_M(r) \bigl(y'(r)\bigr)^2+{\nu \over r^4 f_M(r)}} \,, 
}
where $\CT = N_f\mu_8 R^9 \Omega_4/g_s$, $\mu_8 = (2\pi)^{-8}\ell_s^{-9}$ 
and $'$ denotes the derivative with respect to $r$. 
We calculate the equation of motion from the action \conActzero, 
\eqn\conEOMzero{
{d \over dr}\Biggl[r^2 f_M(r) y'(r) \biggl(r^2f_M(r) \bigl(y'(r)\bigr)^2+{\nu \over r^4 f_M(r)}\biggr)^{-{1/2}} \Biggr] = 0 \,.
}
Integrating this equation with respect to $r$ and 
imposing the boundary condition,
\eqn\conBdry{
y(r_0) = 0 \,, \quad y'(r_0) = \infty\,, 
}
we obtain 
\eqn\conSolzero{
\bigl(y'(r)\bigr)^2 = {\nu \over r^6 \bigl(f_M(r)\bigr)^2}\biggl({r^2f_M(r) \over r_0^2 f_M(r_0)} -1 \biggr)^{-1} \,.
}
Therefore we can describe the separation between the D8-branes 
at the UV cutoff, $r = r_\infty$, as 
\eqn\conSepzero{
l = 2\int_{r_0}^{r_\infty} dr\, y'(r) 
	= 2\int_{r_0}^{r_\infty} dr\, {\nu \over r^3 f_M(r)}\biggl({r^2f_M(r) \over r_0^2 f_M(r_0)} -1 \biggr)^{-{1 \over 2}} \,.
}

Now let us turn on the baryons which are homogeneously distributed 
in $\Bbb{R}^3(x^1,x^2,x^3)$. 
These baryons are realized by baryon vertices, that is, 
the (baryonic) D4-branes wrapping $S^4$ \Wii. 
We assume that the baryon vertices are located on the D8-branes at $r = r_c$. 
From the viewpoint of the worldvolume theory on the D8-branes, 
the number of baryons corresponds to $U(1)$ charge, 
because the Chern-Simons action includes the term, 
$A^{U(1)}_0 \tr (F^{SU(N_f)})^2$. 

More concretely we study the action  of D8-branes. 
It consists of the DBI action and the Chern-Simons action, 
$$
S_8 = S_{\rm DBI} +S_{\rm CS} \,.
$$
The $U(N_f)$ worldvolume gauge field, $\CA$, on the D8-branes 
can be decomposed as 
$$
\CA = A^{SU(N_f)} + {1 \over \sqrt{2N_f}}A^{U(1)} \,.
$$ 
Since we are interested in the U(1) baryon charge, 
we focus on the time component of $U(1)$ gauge field, $A^{U(1)}_0$, 
which depends only on $r$, 
and we assume that the other components are equal to zero. 
For later convenience we define the dimensionless field by 
\eqn\defUoneg{
a(r) := {2\pi\ell_s^2 \over \sqrt{2N_f}R^2} A^{U(1)}_0 \,.
}
Due to this gauge field, the DBI action \conActzero\ is modified as 
\eqn\conDBI{
S_{\rm DBI} = -\CT \int d^4x dr\, \CL_{\rm DBI} \,, \quad 
\CL_{\rm DBI} = \nu \sqrt{r^2f_M(r) \bigl(y'(r)\bigr)^2+{\nu \over r^4 f_M(r)} -{1 \over r} \bigl(a'(r)\bigr)^2} \,.
}

We introduce the baryon vertices at $r=r_c$. 
We then have to take into account the Chern-Simons term 
coupling with the four-form flux $G_{(4)}$, 
\eqn\CSGfour{
{\mu_8 (2\pi \ell_s^2)^3 \over 3!} \int G_{(4)} \wedge \omega_{(5)} = {N_4 \over 24\pi^2} \int \omega_{(5)} \,, \quad 
\omega_{(5)} = \Tr\biggl(\CA\CF^2 -{1 \over 2}\CA^3\CF +{1 \over 10}\CA^5\biggr) \,,
}
because this includes the term inducing the baryon charge, 
\eqn\srcAFF{
S_{\rm CS} = {N_4 N_f \over 24\pi^2} \int d^4x dr {3 \over \sqrt{2N_f}} A^{U(1)}_0 \tr \bigl(F^{SU(N_f)}\bigr)^2 \,. 
}
Since the baryon vertices are at $r=r_c$ and 
are homogeneously distributed in $\Bbb{R}^3$,
$\tr F^2$ is identified 
with the baryon number density as  
\eqn\trFFbarnum{
{1 \over 8\pi^2}\tr \bigl(F^{SU(N_f)}\bigr)^2 = n_B\delta(r-r_c) \,. 
}
Therefore \srcAFF\ becomes  
\eqn\conCS{
S_{\rm CS}=\CT \int d^4xdr\, 3 N_B a(r) \delta(r-r_c) \,, \quad 
N_B := n_B \biggl({2\pi\ell_s \over R} \biggr)^4 \,.
}
We note that, although there is the other RR flux, $G_{(8)}$, in \CONFRRfluxdl, 
it does not couple with $a(r)$, the time component of $U(1)$ gauge field.\foot{Compared with the source of baryon charge \CSGfour, 
$(2\pi\ell_s^2)^3G_{(4)}\wedge A_0^{U(1)} \wedge \CF^2 \sim \CO(\ell_s^9)$, 
we can neglect $(2\pi\ell_s^2)^4 *G_{(8)}\wedge A_0^{U(1)} \wedge \CF^3 \sim \CO(\ell_s^{11})$ 
due to small $\ell_s$. (See also \CONFRRfluxdl.)}

\subsec{Force balance condition between D8-branes and baryonic D4-branes}

Since the baryonic D4-branes are attached on the D8-branes at $r=r_c$, 
the force from D4-branes has to balance with that from the D8-branes \BLL. 

We remind the total action of D8-branes given by \conDBI\ and \conCS, {\it i.e.}, 
\eqna\conACT
$$\eqalignno{
&S_8 = \CT \int d^4x dr 
	\bigl[-\CL_{\rm DBI} +3 N_B a(r) \delta(r-r_c) \bigr] \,, &\conACT{a} \cr 
&\CL_{\rm DBI} = \nu \sqrt{r^2f_M(r) \bigl(y'(r)\bigr)^2+{\nu \over r^4 f_M(r)} -{1 \over r} \bigl(a'(r)\bigr)^2} \,. &\conACT{b}
}$$
The equations of motion for $y(r)$ and $a(r)$ are respectively 
\eqna\conEOM
$$\eqalignno{
&{d \over dr}\Biggl[{r^2f_M(r)y'(r) \over \CL_{\rm DBI}}\Biggr] = 0 \,, &\conEOM{a} \cr
&d'(r) = 3 N_B \delta(r-r_c) \,, &\conEOM{b}
}$$
where we introduced the electric displacement field, 
\eqn\conEDF{
d(r) := -{\delta \CL_{\rm DBI} \over \delta a'(r)} 
= {\nu^2a'(r) \over r\CL_{\rm DBI}} \,.
}
From \conEOM{b} we obtain
$$
d(r) = 3 N_B \,, 
$$
that is, the electric displacement field becomes the constant 
proportional to the baryon density. 
With the boundary condition, $y'(r_0) = \infty$, 
we can solve \conEOM{a} as 
\eqn\conSoly{
\bigl(y'(r)\bigr)^2 ={\nu \over r^6 \bigl(f_M(r)\bigr)^2}\biggl({r^2 f_M(r)(\nu^2 +rd^2) \over r_0^2 f_M(r_0)(\nu^2 +r_0 d^2)} -1\biggr)^{-1} \,. 
}

We are now considering the baryonic D4-branes which are attached 
on the D8-branes at $r=r_c$. 
Then we can imagine two cases about the shape of D8-branes. 
These depend on the direction of force from the baryonic D4-branes.
\bigskip\vbox{
\centerline{\epsfbox{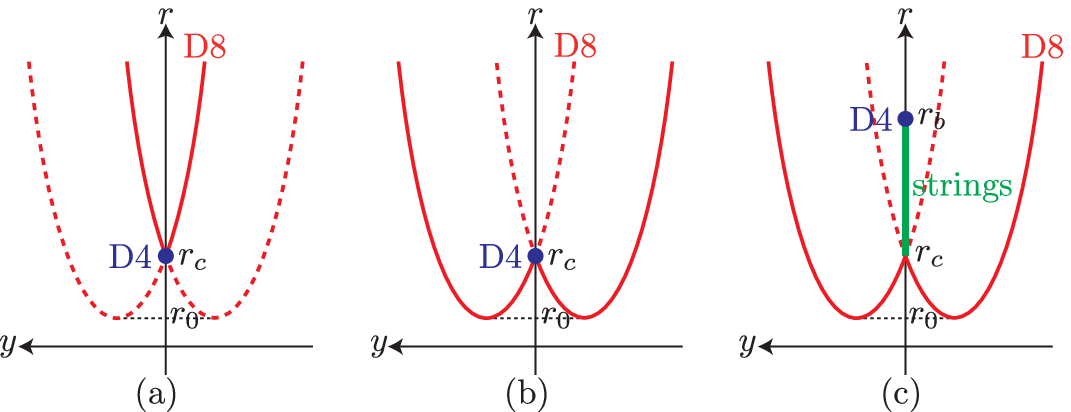}} \smallskip
\centerline{\vbox{\offinterlineskip 
\halign{ \strut# & #\hfil \cr
\fig\figFBC{} & The force balance between the D8-branes and baryonic D4-branes. \cr
& (a) The V-shape D8-branes. (b,c) The W-shape D8-branes.  \cr
}}}
}\bigskip\nobreak\noindent
One is the case as \figFBC a where the V-shape D8-branes are pulled down 
by the falling D4-branes. 
We remember that this is the case for the generalized Sakai-Sugimoto model, 
{\it i.e.}, the D4-branes background without the D0-branes \BLL. 
The other is as \figFBC b where the W-shape D8-branes are hung up 
by the floating D4-branes. 

In order to clarify the shape of D8-branes in our D4+D0 model, 
we calculate the on-shell action of baryonic D4-branes. 
The baryon vertices with density $n_B$ are identified 
with the $n_B v_3$ D4-branes wrapping $S^4$. 
Note that $v_3\, (:= \int d^3x)$ is the volume of $\Bbb{R}^3$. 
Therefore the on-shell action of baryonic D4-branes is 
\eqn\conDfouract{
S_4 = -(n_B v_3) \mu_4 \int dx^0 d\Omega_4\,e^{-\phi}\sqrt{-\det g_{\rm D4}}\biggr|_{\hbox{\sevenrm on-shell}} =: -\int dx^0\, \CE_4 \,, \quad  
\CE_4 = {1 \over 3}\CT v_3 d\sqrt{{\nu \over r_c}} \,. 
}
Since the energy, $\CE_4$, decreases as $r_c$, 
the D4-branes yield an upward force (to the positive direction of $r$). 
Therefore our D4+D0 model prefers the W-shape D8-brane than the V-shape one. 
We also need to take into account the possibility as \figFBC c, 
in which the W-shape D8-branes and the baryonic D4-branes are connected 
by fundamental strings and the position of D4-branes is $r_b\, (\geq r_c)$ \SekSo. 
The action of the baryonic D4-branes and fundamental strings are described as 
$$\eqalignno{
S_{4+{\rm st}} &= -(n_B v_3) \biggl[\mu_4 \int dx^0 d\Omega_4\,e^{-\phi}\sqrt{-\det g_{\rm D4}} 
	+{1 \over 2\pi\ell_s^2}\int dx^0 \int_{r_c}^{r_b} dr\sqrt{-\det g_{\rm string}}\biggr]_{\hbox{\sevenrm on-shell}} \cr
&= -\CT v_3 d\sqrt{\nu} \int dx^0 \biggl[{1 \over 3\sqrt{r_b}} 
	+\int_{r_c}^{r_b}{dr \over \sqrt{r^3-r_M^3}}\biggr]
	=: -\int dx^0\, \CE_{4+{\rm st}} \,.
}$$
Since the derivative of the energy $\CE_{4+{\rm st}}$ with respect to $r_b$ 
is always positive, 
$\CE_{4+{\rm st}}$ has a minimum at $r_b = r_c$. 
In other words, the configuration like \figFBC b is favorable 
than that like \figFBC c. 
Finally the W-shape D8-branes on which the baryonic D4-branes directly attached 
(\figFBC b) is the (at least classically) stable configuration . 

For the W-shape D8-branes we calculate the separation 
at the UV cutoff $r = r_\infty$ in terms of \conSoly, 
\eqnn\conSep
$$\eqalignno{
l &= 2\biggl(-\int_{r_c}^{r_0}+\int_{r_0}^{r_\infty}\biggr) dr\, |y'(r)| \cr
&= 2\biggl(\int_{r_0}^{r_c}+\int_{r_0}^{r_\infty}\biggr) dr {\sqrt{\nu} \over r^3 f_M(r)}\biggl({r^2 f_M(r)(\nu^2 +rd^2) \over r_0^2 f_M(r_0)(\nu^2 +r_0 d^2)} -1\biggr)^{-{1 \over 2}} \,. &\conSep 
}$$
\bigskip\vbox{
\centerline{\epsfbox{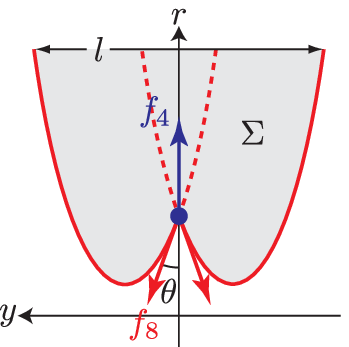}} \smallskip
\centerline{\vbox{\offinterlineskip 
\halign{ \strut# & #\hfil \cr
\fig\figDeDf{} & The force balance condition: $f_8\cos\theta = f_4$. \cr
}}}
}\bigskip\nobreak
Now let us calculate the force balance condition between the D8-branes 
and the baryonic D4-branes (see \figDeDf).\foot{In this paper we calculate the forces in the left half ($y(r) \geq 0$). Therefore the actual total forces are twice and balance as $2f_8\cos\theta = 2f_4$.} 
The Legendre transform of the DBI action of D8-branes \conDBI\ is 
\eqnn\conLegeight
$$\eqalignno{
{\tilde S}_8 &:= \CT \int d^4x dr \Bigl(a'(r) d+\CL_{\rm DBI}\Bigr) \cr
&= \CT \int d^4x dr \sqrt{(\nu^2 +rd^2)\biggl(r^2f_M(r)\bigl(y'(r)\bigr)^2 +{1 \over r^4 f_M(r)}\biggr)} \,. &\conLegeight
}$$
From this equation, we can read that the effective tension at $r=r_c$ is  
\eqn\confeight{
f_8 = {\CT v_3 \over R} \nu^{1 \over 4} \sqrt{{\nu^2 \over r_c} +d^2} \,. 
}
The angle between the D8-branes and the $r$ axis at $r=r_c$ is computed as 
\eqn\conangle{
\cos \theta = {\sqrt{g_{rr}}dr \over \sqrt{g_{rr}dr^2 +g_{yy}dy^2}}\Biggr|_{r = r_c}
= \sqrt{1-{r_0^2 f_M(r_0)(\nu^2 +r_0d^2) \over r_c^2 f_M(r_c)(\nu^2 +r_c d^2)}} \,. 
}
The D8-branes provide the force, $f_8\cos \theta$, along the $r$ direction. 
On the other hand, the force from the D4-branes is evaluated 
in terms of \conDfouract, 
\eqn\conffour{
f_4 = -{1 \over \sqrt{g_{rr}}}{d\CE_4 \over dr_c}\Biggr|_{r=r_c} 
= {\CT v_3 \over 6R} \nu^{1 \over 4} d\sqrt{f_M(r_c)} \,. 
}
As a result, from \confeight, \conangle\ and \conffour, 
we obtain the force balance condition, $f_8\cos\theta = f_4$, namely, 
\eqn\conFBC{
{1 \over 6}d = \sqrt{{\nu^2 +r_c d^2 \over r_c f_M(r_c)}\biggl(1-{r_0^2 f_M(r_0)(\nu^2 +r_0d^2) \over r_c^2 f_M(r_c)(\nu^2 +r_c d^2)}\biggr)} \,. 
}

\subsec{Chiral condensate}

In order to consider the chiral condensate, we adopt the method suggested 
by \AK, in which the open Wilson line operator, 
$$
\CO^j_i(x^\mu) = \psi_L^{\dagger j}(x^\mu,y=-l/2) 
{\cal P}e^{\int_{-l/2}^{l/2} dy (iA_y +\Phi)} \psi_{Ri}(x^\mu,y=l/2) \,, 
$$
was introduced. 
$i,j$ are indices of the fundamental representation of $U(N_f)$. The AdS/CFT correspondence allows us to evaluate the vacuum expectation value 
of the open Wilson line operator by the use of the on-shell action 
of fundamental string whose boundary is the flavor D8-branes (see \figDeDf), 
that is to say, 
$$
\langle \CO_i^j \rangle \simeq \delta_i^j \langle \CO \rangle \,, \quad 
\langle \CO \rangle = e^{-S_\CO} \,, \quad 
S_\CO ={1 \over 2\pi\ell_s^2}\int_\Sigma d^2\sigma \sqrt{\det g} \,, 
$$ 
to leading order of $\alpha'\, (=\ell_s^2)$. 
One can realize $\langle \CO \rangle$ 
as the chiral condensate $\langle{\bar \psi}\psi\rangle$. 
$S_\CO$ is, in other words, the area of the worldsheet instanton. 
By the use of \conSep, $S_\CO$ can be written down as 
\eqn\conOWLarea{
S_\CO = {R^2 \over \pi\ell_s^2}\int_0^{l/2}dy \int_{r(y)}^{r_\infty} dr 
= {R^2 \over \pi\ell_s^2} \biggl(\int_{r_0}^{r_c}+\int_{r_0}^{r_\infty}\biggr) 
	dr (r_\infty -r)|y'(r)| \,.
}
We note that $S_\CO$ does not have UV divergence,  
because we have introduced the UV cutoff, $r_\infty$, in our model 
as we have discussed in Section 2. 
However $S_\CO$ still has a large value depending on the cutoff. 
So we consider $\langle \CO \rangle$ divided 
by the large constant, 
\eqn\conRegu{
\langle \CO \rangle_\infty := \exp \biggl[-{R^2 \over 2\pi\ell_s^2} l(r_\infty -r_M)\biggr] \,.
}

Let us evaluate the chiral condensate numerically. 
In order to sweep out $r_M$ from numerical computations, 
we rescale the variables as 
$$
w:={r \over r_M} \,, \quad {\hat d}:= \sqrt{r_M} d \,, \quad 
{\hat l} := r_M^2 l \,.
$$
Then the force balance condition \conFBC\ is rewritten as  
$$
w_0^2 (1-w_0^{-3})(\nu^2+w_0{\hat d}^2)
= w_c^2 (1-w_c^{-3})\biggl(\nu^2 +w_c {\hat d}^2{35+w_c^{-3} \over 36}\biggr) 
=: Q(\nu,{\hat d},w_c) \,. 
$$
This equation determines $w_0$ as a function of $\nu$, ${\hat d}$ and $w_c$. 
From \conSep, \conOWLarea\ and \conRegu, 
the separation and chiral condensate are described as 
\eqnn\conSepK
\eqnn\conregCCK
$$\eqalignno{
&{\hat l} =  2\biggl(\int_{w_0}^{w_c}+\int_{w_0}^{w_\infty}\biggr) dw {\sqrt{\nu} \over w^3-1}\biggl({w^2 (1-w^{-3})(\nu^2 +w{\hat d}^2) \over  Q(\nu,{\hat d},w_c)} -1\biggr)^{-{1 \over 2}} \,, &\conSepK \cr 
&{\langle \CO \rangle \over \langle \CO \rangle_\infty}  
= \exp \biggl({R^2 \over \pi \ell_s^2 r_M}I\biggr) \,, \cr
&\quad I = \biggl(\int_{w_0}^{w_c}+\int_{w_0}^{w_\infty}\biggr) dw {\sqrt{\nu}  \over w^2+w+1}\biggl({w^2 (1-w^{-3})(\nu^2 +w{\hat d}^2) \over  Q(\nu,{\hat d},w_c)} -1\biggr)^{-{1 \over 2}} \,. &\conregCCK
}$$
By fixing $\nu$, ${\hat l}$ and $w_\infty$, 
we calculate the density ${\hat d}$ dependence of $\exp(I)$. 
\bigskip\vbox{
\centerline{\epsfbox{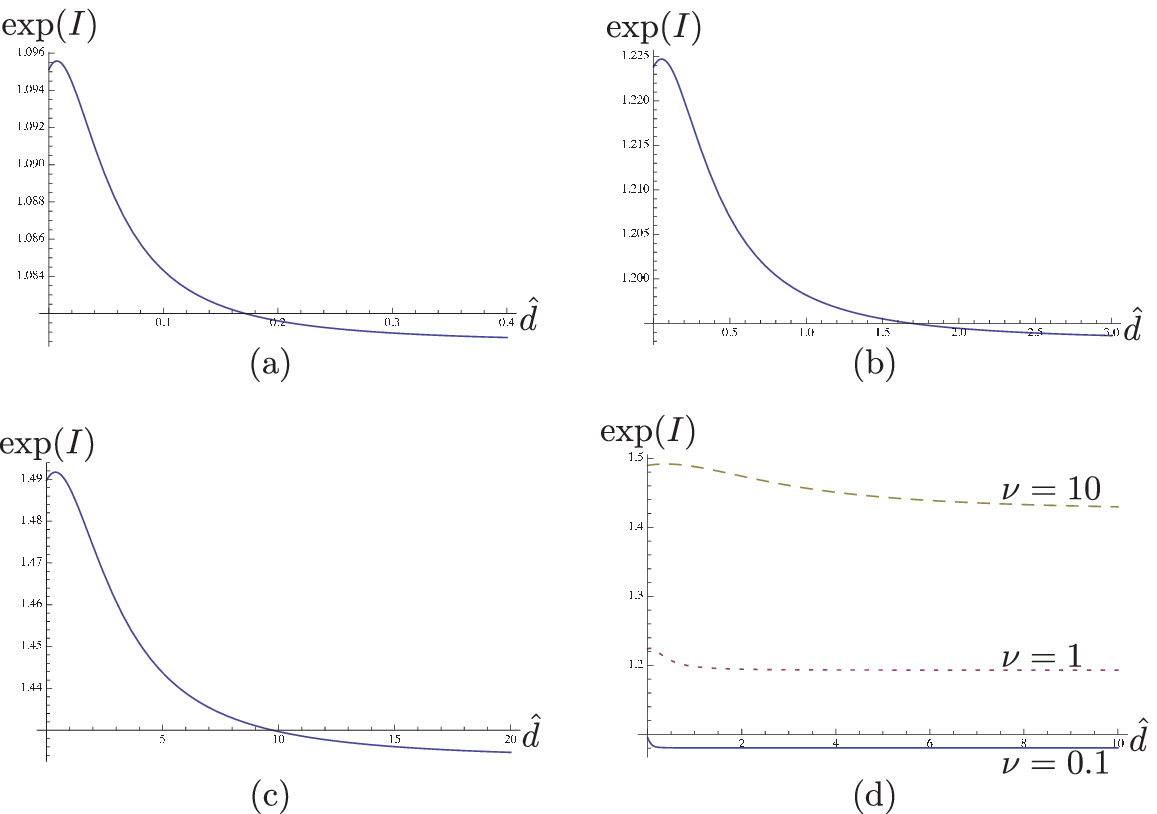}} \smallskip
\centerline{\vbox{\offinterlineskip 
\halign{ \strut# & #\hfil \cr
\fig\figconCC{} & The plots of the density ${\hat d}$ dependence of $\exp(I)$ 
with $w_\infty =10^4$ and ${\hat l}=0.1$ fixed. \cr
&(a) $\nu = 0.1$. (b) $\nu = 1$. (c) $\nu=10$. \cr
&(d) The plots for $\nu$=0.1 (solid), 1 (dotted), 10 (dashed). Every $\exp(I)$ converges to \cr 
&\phantom{(d) }a certain constant at ${\hat d}\to \infty$. \cr
}}}
}\bigskip\nobreak\noindent

For instance, when we set $w_\infty=10^4$, ${\hat l} = 0.1$ and $\nu = 0.1,1,10$, 
we obtain \figconCC.
Note that the contribution from large $w$ 
to the integrations in \conSepK\ and \conregCCK\  
is very small, 
therefore $\exp(I)$ is not sensitive to the cutoff $w_\infty$. 
We can see in \figconCC\ that 
the chiral condensate, $\langle \CO \rangle\, (=\langle \CO \rangle_\infty(e^I)^{R^2/(\pi\ell_s^2r_M)})$, increases in very low density. 
This is different from our intuition from ordinary QCD. 
On the other hand, in high density, the chiral condensate is a monotonically 
decreasing function of ${\hat d}$ 
and converges to a finite value as ${\hat d} \to \infty$. 
In all other known models of holographic QCD \refs{\NSSY\SeoSi\EV{--}\SekSi}, 
the chiral condensate increases in high density and 
it is completely opposite to the expectation from ordinary QCD. 
Therefore one can say that our model drastically improves the high density behavior
of chiral condensate.

\newsec{Comments on the deconfined phase} 

\subsec{The action of flavor D8-branes}

The background corresponding to the deconfined phase has been given 
by \DECmetricdl{} and \DECRRfluxdl. 
We embed the $N_f$ D8-branes (and $N_f$ anti-D8-branes) into it 
in the same manner as we did in the confined phase, that is to say, 
we identify the worldvolume coordinates of the D8-branes 
with $(x^0,x^1,x^2,x^3,r,\Omega_4)$ and 
assume that the collective coordinate, $y$, depends only on $r$. 
Since we are interested in the $U(1)$ baryon charge, 
we take into account the $U(1)$ gauge field denoted by \defUoneg. 
Then the DBI action is described as 
\eqn\decDBI{
S_{\rm DBI}^{\rm (dec)} = -\CT \int d^4x dr\, \CL_{\rm DBI}^{\rm (dec)} \,, \quad 
\CL_{\rm DBI}^{\rm (dec)} = \nu \sqrt{r^2 f_T(r) \bigl(y'(r)\bigr)^2 
	-{1 \over r}\bigl(a'(r)\bigr)^2 +{1 \over r}} \,.  
}

There are two Chern-Simons terms coupling with the $U(1)$ gauge field, $a(r)$.
One is the coupling with the four-form flux \DECRRfluxdl, {\it i.e.}, 
$\int G_{(4)} \wedge \omega_{(5)}$, originated from the background D4-branes. 
As we explained in Section 3.1, $\omega_{(5)}$ includes 
$A^{U(1)}\tr (F^{SU(N_f)})^2$. 
Therefore it provides us the source term of 
the $U(1)$ baryon charge, 
\eqn\decCSb{
S_{\rm CS}^{(B)}=\CT \int d^4xdr\, 3 N_B a(r) \delta(r-r_c) \,,
}
in the same way as \conCS. 
This also implies that the $N_B$ baryonic D4-branes wrapping $S^4$ 
are located at $r = r_c$.  

In the deconfined phase there is the other Chern-Simons term 
induced by the eight-form flux \DECRRfluxdl, $G_{(8)}$, of the D0-branes, 
namely, $\int G_{(8)} \wedge \CA_0 dx^0$. 
This term includes 
\eqn\decCSdzero{
S_{\rm CS}^{(D0)} = \CT \int d^4x dr\, 3\nu y'(r) a(r) \,.
}
We note that this Chern-Simons term yields the $U(1)$ charge 
even if there is not a baryon ($N_B=0$).

\subsec{Without a baryon}

In this subsection we concentrate on the case of no baryon, that is, $N_B = 0$.  
From \decDBI\ and \decCSdzero, the action of D8-branes is 
\eqn\decactionNoB{
S_8^{\rm (dec)}\bigr|_{N_B=0} = -\CT \int d^4xdr \bigl[\CL_{\rm DBI}^{\rm (dec)} 
	-3\nu y'(r) a(r) \bigr] \,. 
}
We calculate the equations of motion from this action, 
\eqna\decEOMNoB
$$\eqalignno{
&{d \over dr}\biggl[{\nu r^2 f_T(r) y'(r) \over \CL_{\rm DBI}^{\rm (dec)}} -3a(r)\biggr] = 0 \,, &\decEOMNoB{a} \cr
&{d \over dr}\biggl[{\nu a'(r) \over r \CL_{\rm DBI}^{\rm (dec)}} -3y(r) \biggr] = 0 \,. &\decEOMNoB{b}
}$$
Since these equations are symmetric under the constant shifts of the fields, 
we redefined the fields by 
\eqn\shiftedfield{
\psi(r) := y(r) +C_1 \,, \quad 
\alpha(r) := a(r) +C_2 \,.
}
Then, integrating \decEOMNoB{} with respect to $r$, we obtain 
\eqna\decEOMNoBre
$$\eqalignno{
&\psi'(r) = 
	{3\alpha(r) \over \sqrt{r^3 f_T(r) \Bigl\{ 9r^3f_T(r)\bigl(\psi(r)\bigr)^2 -9\bigl(\alpha(r)\bigr)^2 +r^2f_T(r) \Bigr\}}} \,, &\decEOMNoBre{a} \cr
&\alpha'(r) = 
	{3 \sqrt{r^3 f_T(r)} \psi(r) \over \sqrt{9r^3f_T(r)\bigl(\psi(r)\bigr)^2 -9\bigl(\alpha(r)\bigr)^2 +r^2f_T(r)} } \,, &\decEOMNoBre{b}
}$$
where the integration constants are absorbed by $C_1$ and $C_2$. 
Note that, when $(\psi(r), \alpha(r))$ is a solution of \decEOMNoBre{}, 
$(-\psi(r), -\alpha(r))$ is also a solution. 

The trivial solution is $\alpha(r) = \psi(r) = 0$. 
When we put a pair of boundary conditions, $y(r_\infty)=\pm l/2$, 
this solution leads to a pair of solutions, $y(r) =\pm l/2$, 
which represents the parallel 
D8-branes and anti-D8-branes separated from each other by $l$. 
This configuration preserves the chiral symmetry, $U(N_f) \times U(N_f)$.

Let us look for a non-trivial solution. 
In order to find a U-shape solution whose tip is located at $r=r_0$, 
we impose the boundary condition:
\eqn\BDCpsipsip{
\psi(r_0) = 0 \,, \quad \psi'(r_0) =\infty \,.
}
From \decEOMNoBre{a}, one can interpret $\psi'(r_0) =\infty$ to  
$r_0^2 f(r_0) -9\bigl(\alpha(r_0)\bigr)^2 = 0$.  
Therefore the boundary condition \BDCpsipsip\ can be replaced with 
\eqn\BDCpsialpha{
\psi(r_0) = 0\,, \quad \alpha(r_0) = {1 \over 3}r_0 \sqrt{f(r_0)} \,.
}
We are now considering the $N_B = 0$ case, {\it i.e.}, 
there is no source induced by a baryon.
Therefore there does not exit a cusp singularity on the D8-branes. 
In terms of the solution of \decEOMNoBre{} and \BDCpsialpha, 
such a smooth U-shape D8-brane is described as 
\eqn\decUsol{
y(r) = \pm \psi(r) \,, \quad 
a(r) = \pm(\alpha(r) - \alpha(r_0)) \,, 
}
where $C_1$ and $C_2$ in \shiftedfield\ are determined 
so that $y(r)$ and $a(r(y))$ are smooth at $y(r_0) = 0$. 
\bigskip\vbox{
\centerline{\epsfbox{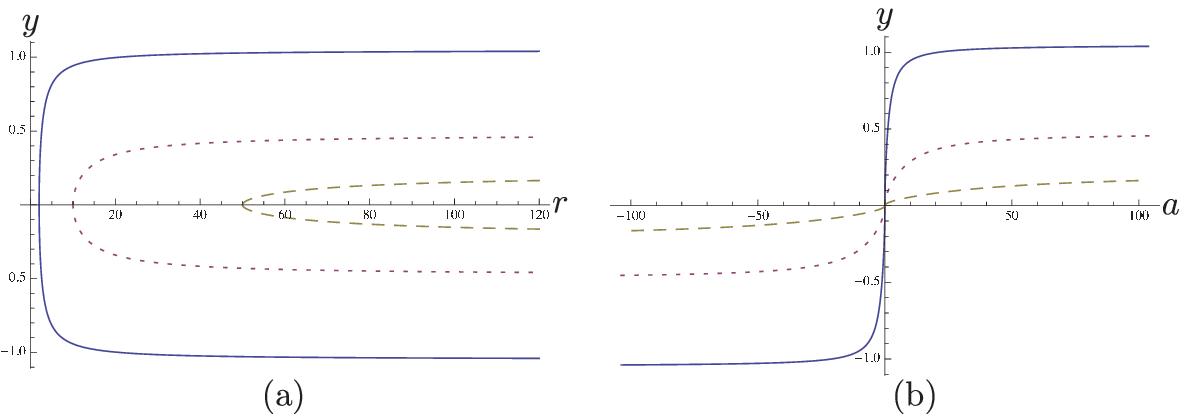}} \smallskip
\centerline{\vbox{\offinterlineskip 
\halign{ \strut# & #\hfil \cr
\fig\figdUsol{} & The plots of $y(r)$ and $a(r)$ with fixed $r_T=1$ for $r_0$= 2 (solid), 10 (dotted) and \cr
& 50 (dashed). (a) The plots of $y(r)$. (b) The parametric plots of $(a(r), y(r))$. \cr
}}}
}\bigskip\nobreak\noindent
Since it is difficult to solve \decEOMNoBre{} analytically, 
we perform numerical computation. 
For instance, in fixing $r_T=1$, 
we obtain the solutions depicted in \figdUsol. 
The D8-branes and anti-D8-branes are smoothly connected at $r=r_0$, 
so that the U-shape D8-branes appear.  
This implies the chiral symmetry breaking from $U(N_f) \times U(N_f)$ 
to $U(N_f)$. 
Although the gauge field diverges at $r \to \infty$, 
we neglect the large $r$ region. 
Because, as we discussed in Section 2.2, the UV cutoff $r_\infty$ is 
necessary due to the validity of supergravity approximation.

\subsec{With baryons}

We shall turn on the baryons which are the $n_B v_3$ baryonic D4-branes 
located at $r=r_c$. 
The total action of D8-branes is given by \decDBI, \decCSb\ and \decCSdzero\ 
because of $N_B \neq 0$. 
Although the equation of motion for $y(r)$ is same as \decEOMNoB{a},  
the one for $a(r)$ is slightly different from \decEOMNoB{b} due to 
the Chern-Simons term \decCSb. 
By redefining  
\eqn\decEDF{
d(r) := -{\delta \CL_{\rm DBI}^{\rm (dec)} \over \delta a'(r)} -3\nu y(r) \,,  
}
the equation of motion for $a(r)$ has a simple expression,  
$d'(r) = 3N_B \delta(r-r_c)$.
Therefore $d(r)$ is a constant equal to $3N_B$.

In terms of \DECmetricdl{} we write down the on-shell action of 
the baryonic D4-branes wrapping $S^4$ as
\eqn\decDfouract{
S_4^{\rm (dec)} = -\int dx^0 \CE_4^{\rm (dec)} \,, \quad 
\CE_4^{\rm (dec)} = {1 \over 3}\CT v_3 d r_c \sqrt{f_T(r_c)} \,. 
}
Since the energy $\CE_4^{\rm (dec)}$ monotonically increases as $r_c$, 
the baryonic D4-branes obtain a force toward the negative direction of $r$. 
Therefore this force from the D4-branes balances with the force 
from the V-shape D8-branes in a manner as \figFBC a. 
By using the solution of $\psi(r)$ and $\alpha(r)$ 
that we obtained in the previous subsection, 
we can describe the V-shape D8-brane as 
\eqn\decVsol{
y(r) = \pm(\psi(r) - \psi(r_c)) \,, \quad 
a(r) = \pm(\alpha(r) - \alpha(r_0)) \,. 
}
Note that the gauge field $a(r)$ is not continuous at $r = r_c$ 
because there is the baryonic D4-branes, namely, the source of $U(1)$ charge. 

The Legendre transformed action of D8-branes is
\eqn\decLegeight{
{\tilde S}_8^{\rm (dec)} = \CT \int d^4x dr \sqrt{\bigl(\nu^2 +r(d+3\nu y(r))^2 \bigr)
	\biggl(r^2 f_T(r) \bigl(y'(r)\bigr)^2 +{1 \over r}\biggr)} \,, 
}
up to total derivative. 
From \decDfouract\ and \decLegeight\ we calculate the force balance condition, 
\eqn\decFBC{
{1 \over 3}d r_c \Biggl[ 1+{1 \over 2}\biggl({r_T \over r_c}\biggr)^3 \Biggr] 
= \sqrt{f_T(r_c)(\nu^2 +r_c d^2) \over r_c^2 f_T(r_c) \bigl(y'(r_c)\bigr)^2 + r_c^{-1}} \,.
}
The separation of V-shape D8-branes at $r=r_\infty$ is 
\eqn\decSep{
l = 2\int_{r_c}^{r_\infty} |y'(r)| = 2(\psi(r_\infty) -\psi(r_c)) \,. 
}
In the same way as Section 3.3, we shall study the density dependence of 
the chiral condensate $\langle \CO\rangle$ by 
\eqnn\decCCK
$$\eqalignno{
{\langle \CO\rangle \over \langle \CO\rangle_\infty} 
	&= \exp \biggl[{R^2 \over \pi\ell_s^2}\int_{r_c}^{r_\infty}dr \int_0^{y(r)}dy {\sqrt{\nu} \over r^3 f_T(r)}
	-{R^2 \over \pi\ell_s^2}\int_{r_c}^{r_\infty}dr \int_0^{l/2}dy {\sqrt{\nu} \over r^3 f_T(r)} \biggr] \cr 
	&= \exp \biggl({R^2\sqrt{\nu} \over 2\pi\ell_s^2} I_{(D)}\biggr) \,, \quad 
I_{(D)} = \int_{r_c}^{r_\infty} dr {2y(r) -l \over \sqrt{r^3 -r_T^3}} \,. &\decCCK 
}$$
\bigskip\vbox{
\centerline{\epsfbox{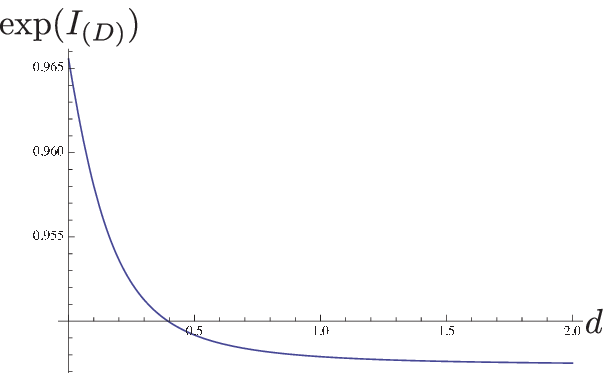}} \smallskip
\centerline{\vbox{\offinterlineskip 
\halign{ \strut# & #\hfil \cr
\fig\figdecCC{} & The plot of $\exp(I_{(D)})$ depending on the density $d$ with fixed $l=0.5$ \cr
& and $r_T=\nu = 1$. \cr
}}}
}\bigskip\nobreak\noindent

For instance we numerically evaluate $\exp(I_{(D)})$ 
with fixed $l=0.5$ and $r_T=\nu = 1$, 
which is depicted in \figdecCC. 
It shows that the chiral condensate monotonically decreases as the density $d$. 
This agrees very well with our expectation from 
the ordinary QCD. 
The $\exp(I_{(D)})$ for $l=0.5$ and $r_T=\nu = 1$ is approximately equal to 0.966 
at $d=0$, and converges to a finite value,  $0.947$, at $d =\infty$ (see \figdecCC).

\newsec{Conclusions}

In this paper we proposed a model of holographic QCD
whose background has  the asymptotic freedom. 
It is the specific field theory limit of the supergravity solution of 
the color D4-branes on which the D0-branes are smeared. 
In taking the limit, we fixed the ratio $\nu$ to be finite. 
Into this background we have introduced the flavor D8-branes and 
the baryonic D4-branes in order to study the chiral condensate 
in dense baryonic medium.
The dilaton in the background of D4-branes without D0-branes 
behaves like $e^\phi \sim r^{3/4}$. 
On the other hand, 
the dilaton in our background behaves like $e^\phi\sim r^{-3/2}$, 
that is, the effective coupling is asymptotically free. 
Although it is still different from $(\log r)^{-1}$ in ordinary QCD, 
the D0-branes improve the asymptotic freedom in our model.

We have studied the chiral condensate by   use of the open Wilson line, 
which we can evaluate by the area of open string worldsheet 
whose boundary is the D8-branes. 
In the confined phase the baryonic D4-branes balance with 
the D8-branes which  have  W-shape rather than V-shape. 
This is because the force from the D4-branes is toward the positive direction 
of the radial coordinate. This can be understood that the curvature increases 
for large radius so that the gravity is acting on baryon vertex upward. 

Under this force balance condition we have calculated 
the chiral condensate concretely with the fixed ${\hat l}$, the separation 
between the D8-branes at the UV cutoff. 
The results for three cases of $\nu$ are shown in \figconCC. 
All of the three have the common features as follows. 
In the region of very low density of baryons, 
the chiral condensate increases as the density. 
This behavior  is not what one expects from the    QCD. 
We attribute this anomaly to the approximation of the uniform distribution of baryons 
in the space $\Bbb{R}^3$, which is not a good one when the density of baryon is small. 
In high density we have obtained a remarkable result, that is, 
the chiral condensate monotonically decreases. 
This is in good agreement with our intuition from the ordinary QCD. 
When the density goes to infinity, the chiral condensate 
converges to a certain constant. 

We have studied also the deconfined phase. 
We numerically solved the equations of motion without the baryons, 
and obtained the U-shape solution of D8-branes. 
It seems that the solution of the gauge field has UV divergence. 
However we do not need to seriously worry about the divergence, 
because basically the UV cutoff is necessary in our background 
due to the validity of supergravity approximation of bulk theory. 
Turning on baryons, we have calculated the chiral condensate. 
It is different from the confined phase that 
the D8-brane balancing with the baryonic D4-branes has V-shape 
in the deconfined phase. 
As a result, it monotonically decreases in whole range 
of the density (see \figdecCC). 
This is  in agreement with the expectation from ordinary QCD. 
In the deconfined phase there is the solution of parallel D8-branes, 
which preserves the chiral symmetry. 
Therefore we expect the chiral symmetry restoration occurs in 
certain high density \ASY. 
One can guess that this transition is clarified 
by comparing the free energies of two configurations, namely, 
the V-shape D8-branes with baryon vertices and the parallel D8-branes. 
However there is subtlety due to the UV cutoff. 
Therefore we need to develop a way beyond the UV cutoff.

\bigbreak\bigskip\bigskip\centerline{{\bf Acknowledgement}}\nobreak

This work was supported by the NRF grant through the SRC program CQUeST 
with grant number 2005-0049409
and partially supported by the WCU project of Korean Ministry 
of Education, Science and Technology (R33-2008-000-10087-0).

\listrefs
\bye